\RequirePackage{fix-cm}
\documentclass{svjour3}                     
\smartqed  
\usepackage{graphicx}
\usepackage{amssymb}

\begin{document}

\title{Cryptanalysis of a multi-party quantum key agreement
protocol with single particles
}


\author{Wei Huang   \and
         Qiao-Yan  Wen  \and
         Bin Liu \and
         Qi Su \and
         Fei Gao
}

\institute{W. Huang \and
Q-Y Wen  \and
        B Liu  \and
        Q Su \and
        F Gao
        \at
State key Laboratory of Networking and Switching Technology, \\Beijing University of Posts and Telecommunications, Beijing 100876 China\\
              \email{huangwei096505@yahoo.cn}
              \and
              W. Huang  \at
State Key Laboratory of Information Security,\\Institute of Software, Chinese Academy of Sciences, Beijing 100190, China\\
}
\date{Received: date / Accepted: date}

\maketitle

\begin{abstract}
Recently, Sun et al. [Quant Inf Proc DOI: 10.1007/s11128-013-0569-x] presented an efficient multi-party quantum key agreement (QKA) protocol by employing single particles and unitary operations. The aim of this protocol is to fairly and securely negotiate a secret session key among $N$ parties with a high qubit efficiency. In addition, the authors claimed that no participant can learn anything more than his/her prescribed output in this protocol, i.e., the sub-secret keys of the participants can be kept secret during the protocol. However, here we points out that the sub-secret of a participant in Sun et al.'s protocol can be eavesdropped by the two participants next to him/her. In addition, a certain number of dishonest participants can fully determine the final shared key in this protocol. Finally, we discuss the factors that should be considered when designing a really fair and secure QKA protocol.

\keywords{Quantum cryptography \and Quantum key agreement  \and Cryptanalysis}
\end{abstract}

\section{Introduction}
\label{intro}

Key agreement (KA) is one of the most basic cryptographic primitives which allows two or more participants to establish a common secret key fairly based on their exchanged information. In contrast to key distribution (KD), in which only one participant determine the secret key and then distributes it to the others, each participant in a KA protocol should contribute his/her influence to the shared key. In other words, the shared key cannot be determined by any non-trivial subset of the participants involved in a QKA protocol. In 1976, Diffe and Hellman \cite{01_DH} first introduced a secure and fair protocol for two parties to agree on a shared key. Since the pioneering work of Diffe and Hellman, much attention has been focused on extending the two-party Diffie-Hellman protocol to the multi-party setting, and a number of correlated protocols have been proposed \cite{02_ITW,03_STM,04_BD}.

However, the security of the classical KA protocols is always based on the assumption of computational complexity. Along with the proposing of efficient algorithms and the development of the computing capability, especially the rapid development of quantum algorithms and quantum computer \cite{05_shor,06_Grover}, classical KA agreement protocols faces more and more serious challenges. Therefore, the development of quantum key agreement (QKA) protocols \cite{07_Zhou,08_Tsai,09_Chong,010_QKA2012,011_LB2012,012_SZW2013}, whose security  only relies on the laws of quantum mechanics (such as quantum no-cloning theorem and Heisenberg uncertainty principle), has become a research hotspot.

In 2004, Zhou et al. presented the first QKA protocol with quantum teleportation technique and maximally entangled states over public channels \cite{07_Zhou}. However, Tsai and Hwang pointed out that a party in this protocol can fully
determine the shared key alone without being detected \cite{07_Zhou}. Hence Zhou et al.'s protocol is not a fair QKA protocol. In 2011, Chong et al. presented a QKA protocol based on the famous BB84 protocol in which the technique of delayed measurement and certain kinds of unitary operations are utilized \cite{09_Chong}. In 2012,  an extension of the two-party quantum key agreement, the first multi-party quantum key agreement (MQKA)  protocol, was proposed by Shi and Zhong by employing EPR pairs and entanglement swapping \cite{010_QKA2012}. Unfortunately, Liu et al. pointed out
that their protocol was not fair as a dishonest participant can totally determine the shared key, and they also presented a secure multi-party QKA protocol only with single particles and single-particle measurements.

It is known that design and cryptanalysis have always been important branches of cryptography. Both of them drive the development of this field. In fact, cryptanalysis is an important and interesting work in quantum cryptography. It estimates the security level of a protocol, finds potential loopholes, and tries to overcome security issues \cite{013-qsj09}. As pointed out by Lo and Ko,\emph{ breaking cryptographic systems was as important as building them}\cite{014-qsj09}. To date, many kinds of attacks strategies have been presented, such as intercept-resend attack \cite{015-GaoPRL}, correlation-extractability (CE) attack \cite{016-GF07,017-qsj10}, Trojan horse attack \cite{018-N.Gisin}, participant attack \cite{019_GF,020-HW13,021_GF08} and so on.

Recently, Sun et al. \cite{012_SZW2013} pointed out that the qubit efficiency of Liu et al.'s MQKA protocol (i.e., $\frac{1}{(k+1)N(N-1)})$ is quite low. To improve the qubit efficiency of the MQKA protocol, they proposed a more efficient one with single particles and unitary operations, the qubit efficiency of which reaches $\frac{1}{(k+1)N}$. For the sake of simplicity, we will call it SMQKA protocol later. The authors of Ref. [12] claimed that the SMQKA protocol satisfies the following four principles.
\begin{itemize}
  \item [$\bullet$]\textbf{Correctness}: Each of the participants involved in this protocol could get the correct shared key.
  \item [$\bullet$]\textbf{Security}:  An outside eavesdropper can get no useful information of the shared key without being detected in the eavesdropping detection.
  \item [$\bullet$]\textbf{Fairness}: All involved participants are entirely peer entities and can equally influence the final shared key. In other words, no non-trivial subset of the participants can determine the shared key.
  \item [$\bullet$]\textbf{Privacy}: No participant can learn anything more than his/her prescribed output in this protocol, i.e., the sub-secret keys of the participants can be kept secret in this protocol.
 \end{itemize}

Unfortunately, we find that the SMQKA protocol cannot achieve privacy and fairness, which indicates that this protocol cannot reach the high efficiency fairly and secretly as Sun et al. claimed. Concretely, the sub-secret of a participant in this protocol can be easily deduced by the two participants next to him/her. More importantly, a certain number of dishonest participants can cooperate to decide the final shared key according to their needs, without being found by the honest participants. The rest of this paper is arranged as follows. In next section, we make a brief introduction of the SMQKA protocol. In Sect. 3, we make an analysis of the SMQKA protocol to show that this protocol can achieve neither privacy nor fairness in detail. Finally, a discussion about the factors that should be considered when designing a really fair and secure QKA protocol, as well as a short conclusion is given in Sect. 4.

\section{Brief review of the SMQKA protocol}
Herein we briefly describe the SMQKA protocol \cite{012_SZW2013} in which $N$ participants are involved. Each participant $P_i$ has a sub-secret key $k_i$, for 0$\leq$$i$$\leq$$N$-1. None of them is willing to divulge any information of his/her sub-secret key to others. This protocol is designed in the travelling mode, which indicates that $P_i$ always sends messages to $P_{i+1}$, where $P_N$=$P_0$. The specific steps of this protocol can be described as follows.
\begin{enumerate}
\item[(1)] Initialization phase.  For each participant $P_i$, he/she prepares a sequence (denoted as $S_i$) of $n$ single particles, each of which is randomly in one of the two polarization
states: $|0\rangle$ and $|1\rangle$. Then he/she generates $kn$ decoy particles which are randomly in one of the four states in \{$|+\rangle$, $|-\rangle$, $|+y\rangle$, $|-y\rangle$\} and inserts them randomly into the sequence $S_i$, where
\begin{eqnarray}
    &&|+\rangle=\frac{1}{\sqrt{2}}(|0\rangle+|1\rangle),\quad\quad\quad\;|-\rangle=\frac{1}{\sqrt{2}}(|0\rangle-|1\rangle),\nonumber\\
    &&|+y\rangle=\frac{1}{\sqrt{2}}(|0\rangle+i|1\rangle),\quad\quad|-y\rangle=\frac{1}{\sqrt{2}}(|0\rangle-i|1\rangle).
\end{eqnarray}
Here $k$ is the detection rate and the new sequence is denoted as $S_i^i$. After that, $P_i$ sends the sequence $S_i^i$ to $P_{i+1}$.
\item[(2)]Eavesdropping detection phase. After the reception of $S_i^i$, $P_{i+1}$ begins to check eavesdropping with $P_i$ as follows. $P_i$ announces the position and the corresponding measuring basis for each of the decoy particles. Then $P_{i+1}$ measures it with the correct basis and inform $P_{i}$ of the measurement outcome. With the measurement outcomes of all the decoy particles, $P_i$ analyzes the security of the transmission of $S_i^i$. If the error rate is higher than a predetermined threshold, they abort the protocol; otherwise, they continue to the next step.
\item[(3)]The message coding phase. Once the eavesdropping detection is finished, $P_{i+1}$ encodes each of the particles in $S_i$ with unitary operation $I$ or $U$ according to his/her sub-secret $k_{i+1}$. Specifically, if a bit in $k_{i+1}$ is 0 (1), he performs the operation $I$ ($U$) on the corresponding particle in $S_i$, where
\begin{eqnarray}
   I=|0\rangle\langle0|+|1\rangle\langle1|,\quad\quad U=i\sigma_y=|0\rangle\langle1|-|1\rangle\langle0|.
\end{eqnarray}
Before sending $S_i$ to the next participant, $P_{i+1}$ also makes use of $kn$ decoy particles to ensure the secure transmission of $S_i$, similar to what $P_i$ does in step (2).  Afterwards, $P_{i+1}$ sends the new sequence ( denoted as $S_i^{i+1}$) to $P_{i+2}$.
\item[(4)] The participants $P_{i+2}$, ..., $P_{i-2}$  execute the eavesdropping detection phase and
 message coding phase in the same way as participant $P_{i+1}$ does in steps (2)-(3) one by one. That is, one after another, they first check eavesdropping, if the transmission
is insecure, the aborts the protocol; otherwise, they encode their sub-secret keys on the particles of $S_i$, and then inserts decoy particles randomly in $S_i$. Finally, they send the new sequence to the next
participant.
\item[(5)] After $P_{i-1}$ receives the sequence $S^{i-2}_i$ sent from $P_{i-2}$, they first checks
eavesdropping with the inserted decoy particles. If there exists no eavesdropping, $P_{i-1}$ encodes his/her sub-secret key $k_{i-1}$ on the particles of $S_i$,  and inserts the $kn$ decoy particles randomly in it, denoted as $S_i^{i-1}$; otherwise, they discard the transmission and abort the protocol.
\item[(6)] Once confirming that each of the participants $P_0$, ..., $P_i$, ..., $P_{N-1}$ has executed the steps (1)-(5), $P_{N-1}$, ..., $P_{i-1}$, ..., $P_{N-2}$ send sequences $S_0^{N-1}$, ..., $S_i^{i-1}$, ..., $S_{N-1}^{N-2}$ to $P_0$, ..., $P_i$, ..., $P_{N-1}$, respectively.
\item[(7)] After $P_i$ receives the sequence $S_i^{i-1}$ sent from $P_{i-1}$, he/she and $P_{i-1}$  check eavesdropping with the decoy particles. If there exists eavesdropping in the quantum channel, they abandon the protocol; otherwise,  $P_i$ measures each of the particles in $S_i$ with the basis $\{|0\rangle, |1\rangle\}$. Since $S_i$ is prepared by $P_i$, he/she knows  the original state of each particle in $S_i$, hence he/she can extract the encoded secret $K^{i-1}_i$=$k_{i+1}$$\oplus$$k_{i+2}$$\oplus$$\cdots$$\oplus$$k_{i-1}$. Finally, $P_i$ obtains the final shared key $K$=$k_i$$\oplus$$K^{i-1}_i$=$k_0$$\oplus$$k_1$$\oplus$$\cdots$$\oplus$$k_{N-1}$, where $i$=0, 1, ..., $N$-1.
\end{enumerate}
\section{Analysis of the SMQKA Protocol}
In this section,  we first show that the SMQKA Protocol cannot achieve the principle of privacy. Then we illustrate that this protocol cannot achieve the principle of fairness, either.
\subsection{The defect on privacy}
Herein we show that the sub-secret key of any involved participant in the SMQKA protocol can be obtained by the two participants next to him/her. Without loss of generality, we consider the situation where $P_{i-1}$ and $P_{i+1}$ try to steal the sub-secret key of $P_i$ (i.e., $k_i$), for 0$\leq$$i$$\leq$$N$-1. To eavesdrop $k_i$, $P_{i-1}$ prepares the sequence $S_{i-1}$ of $n$ single particles which are all in state $|0\rangle$. Then he/she generates $kn$ decoy particles which are randomly in one of the four states in \{$|+\rangle$, $|-\rangle$, $|+y\rangle$, $|-y\rangle$\} and inserts them randomly into $S_{i-1}$. After that, $P_{i-1}$ sends the new sequence (denoted as $S^{i-1}_{i-1}$) to $P_i$. Once $P_i$ receives the sequence $S^{i-1}_{i-1}$, he/she and $P_{i-1}$ check eavesdropping with the decoy particles. Since the decoy particles in $S^{i-1}_{i-1}$ are prepared and inserted by $P_{i-1}$, $P_i$ will find no abnormal occurrence if there quantum channel is secure. Once they confirm there exists no eavesdropping in the transmission of $S^{i-1}_{i-1}$, $P_i$ encodes his sub-secret key $k_i$ on $S_{i-1}$ as described in step (3). Afterwards, $P_i$ also randomly inserts $kn$ decoy particles into $S_{i-1}$ and sends the new sequence (denoted as $S^{i}_{i-1}$) to $P_{i+1}$. Then $P_i$ and $P_{i+1}$ check eavesdropping with the decoy particles inserted by $P_i$. If there exists no eavesdropping in the quantum channel, $P_{i+1}$ can easily deduce $k_i$ as follows. Concretely, he/she measures each of the particles in $S_{i-1}$ with the measuring basis $\{|0\rangle, |1\rangle\}$. If the measurement outcome is $|0\rangle$ ($|1\rangle$), the corresponding key bit in $k_i$ is 0 (1). So far, we have shown that $P_{i-1}$ and $P_{i+1}$ can easily eavesdrop the sub-secret key of $P_i$, for 0$\leq$$i$$\leq$$N$-1. In other words, the SMQKA Protocol cannot achieve the principle of privacy.
\subsection{The defect on privacy}
Now we illustrate that a certain number of dishonest participants can determine the final shared key according to their needs. First, we consider the special circumstance in which $N$-1 dishonest participants try to determine the final shared key. Without loss of generality, we suppose that the $N$-1 dishonest ones are $P_0$, $P_1$, ..., $P_{N-2}$. To determine the final shared key, $P_0$, $P_1$, ..., $P_{N-2}$ pretend to execute the protocol honestly. Specifically, $P_0$, $P_1$, ..., $P_{N-3}$ do nothing on $S_{N-1}$, which is prepared by $P_{N-1}$. Meanwhile, by utilizing the attacking strategy introduced in the previous sub-section, $P_{N-2}$ and $P_0$ steal the sub-secret key of $P_{N-1}$, i.e., $k_{N-1}$. Once the dishonest participants obtain $k_{N-1}$,  they can fully control the final shared key as follow. If their favorite shared key is $k^\prime$, when $P_{N-2}$ has securely received $S_{N-1}$, he encodes the particles of it with $k^\prime$$\oplus$$k_{N-1}$. Concretely, if the $j$-th bit of $k^\prime$$\oplus$$k_{N-1}$ is 0 (1), he/she performs the unitary operation $I$ ($U$) on the corresponding particle in $S_{N-1}$, for 0$\leq$$j$$\leq$$n$. After that, $P_{N-2}$ inserts $kn$ decoy particles in $S_{N-1}$ and sends $S_{N-1}^{N-2}$ to $P_{N-1}$. If the transmission of $S_{N-1}^{N-2}$ is secure, the final shared key obtained by $P_{N-1}$ is $k^\prime$$\oplus$$k_{N-1}$$\oplus$$k_{N-1}$=$k^\prime$. In other words, the $N$-1 dishonest participants have successfully determined the final shared key.

Now we wonder that whether the dishonest participants can determine the final shared key when there exists more than one honest participants. In fact, if the honest participants are nonadjacent, the dishonest participants can fully control the final shared key in the SMQKA protocol. Without loss of generality, we assume that the honest participants are $P_{h_1}$, $P_{h_2}$, ..., $P_{h_s}$, where  $h_i$, $h_j$$\in$\{1, ..., $N$-1\}, $h_i$$\neq$$h_j$$\pm$1. Since the honest participants are nonadjacent, the number of them should be less than half of $N$, i.e., $s$$\leq$$\frac{N}{2}$ ($\frac{N}{2}$-1) if $N$ is even (odd). To control the final shared key, the dishonest participants also pretend to execute the protocol honestly. During the execution of this protocol, they preserve the sequences prepared by the honest ones, i.e., $S_{h_1}$, $S_{h_2}$, ..., $S_{h_s}$. At the same time,  by employing the attacking strategy presented above, the dishonest participants steal the sub-secret keys of the honest participants, i.e., $k_{h_1}$, $k_{h_2}$,..., $k_{h_s}$. Suppose their favorite key is $k^{\prime\prime}$,  they can control the final shared key as follows. For $P_{h_i}$, $P_{h_i-1}$ encodes $S_{h_i}$ with $k^{\prime\prime}$$\oplus$$k_{h_i}$, for $i$=$h_1$, $h_2$, ..., $h_s$. Afterwards, $P_{h_i-1}$ sends $S_{h_i}$ together with $kn$ decoy particles to $P_{h_i}$. If there exists no outside eavesdropping, the final shared key obtained by $P_{h_i}$ is $k^{\prime\prime}$$\oplus$$k_{h_i}$$\oplus$$k_{h_i}$=$k^{\prime\prime}$. Thus far, we have shown that the dishonest participants can fully determine the final shared key in this protocol provided the honest participants are nonadjacent.

\section{Discussion and conclusions}
\subsection{Discussion}
In a real practical quantum key establishing process, including both QKD and QKA, there are two main processes. Here we call these two processes as quantum exchange process and classical postprocessing process, respectively. In the quantum exchange process, the participants make use of quantum states as information carriers to guarantee the security the information transmission based on the principles of quantum mechanics. After this process, the participants can get a sequence of classical string which is usually called raw key. However, due to the noise of the quantum channel and eavesdropping, there always exists certain number of errors in the raw key. In QKD protocols, these errors are usually corrected and cleaned by classical postprocessing process which usually consists of two processes: the information reconciliation process and the privacy amplification process \cite{022_Gisin N,023_Inamori H}.

However, all the existing information reconciliation processes and privacy amplification processes utilized in QKD protocols cannot be directly applied to the QKA protocols, since the dishonest participants may undermine the fairness of the shared key during these two processes. Thus far, all the existing QKA  protocols \cite{07_Zhou,08_Tsai,09_Chong,010_QKA2012,011_LB2012,012_SZW2013} only have concerned the quantum exchange process. In other words, the final shared key established by these protocols are just raw key, which cannot be used to encrypt secret message directly in real life. Obviously, to design a really practical and fair QKA protocol, one should not only consider the fairness in the quantum exchange process, but also present new information reconciliation process and privacy amplification process which can be utilized in QKA protocols for negotiating key fairly. Therefore, how to design the a really unconditional fair and secure QKA protocol,  which involves both the quantum exchange process and classical postprocessing process, still remains an open problem. Some of us are currently investigating this problem and the relevant results will be published in another paper.

\subsection{Conclusion}
In summary, we make an analysis of the MQKA protocol which have been presented recently \cite{012_SZW2013} and point out that this protocol can achieve neither privacy nor fairness as the authors claimed.  Moreover, we make a discussion about the factors that should be taken into consideration when designing a really fair and secure QKA protocol.

\begin{acknowledgements}
This work is supported by NSFC (Grant Nos. 61272057, 61170270, 61100203, 61003286, 61121061), NCET (Grant No. NCET-10-0260), SRFDP (Grant No. 2009000 5110010), Beijing Natural Science Foundation (Grant Nos. 4112040, 4122054), the Fundamental Research Funds for the Central Universities (Grant No. 2011YB01), BUPT Excellent Ph.D. Students Foundation (Grant Nos. CX201217, CX201334).
\end{acknowledgements}


\begin{thebibliography}{}
%

\bibitem{01_DH}Diffie, W.,  Hellman, M.: New directions in cryptography. IEEE Trans. Inf. Theory \textbf{22}, 644-654 (1976)
\bibitem{02_ITW}Ingemarsson, I., Tang, D.T., Wong, C.K.: A conference key distribution system. IEEE Trans. Inf. Theory  \textbf{28}, 714-719 (1982)
\bibitem{03_STM}Steiner, M., Tsudik, G.,  Waidner, M.: Key agreement in dynamic peer groups. IEEE Transactions on Parallel and Distributed Systems \textbf{11}, 769-780 (2000)
\bibitem{04_BD}Burmester, M., Desmedt, Y.: A secure and efficient conference key distribution system. in:
Advances in Cryptology-EUROCRYPT 1994, Lecture Notes in Computer Science, \textbf{950}, 275-286 (1994)
\bibitem{05_shor}Shor, P.W.: Algorithms for quantum computation: Discrete logarithms and factoring. In: Proceedings
of 35th Annual Symposium on Foundations of Computer Science, pp. 124¨C134. Los Alamitos (1994)
\bibitem{06_Grover}Grover, L.K.:Afast quantum mechanical algorithm for database search. In Proceedings of 28thAnnual
ACM Symposium on the Theory of Computing, pp. 212¨C219. Philadelphia (1996)
\bibitem{07_Zhou}Zhou, N., Zeng, G., Xiong, J.: Quantum key agreement protocol. Electron. Lett. \textbf{40}, 1149 (2004)
\bibitem{08_Tsai}Chong, S.K., Tsai, C.W., Hwang, T.: Improvement on Quantum Key Agreement Protocol with Maximally
Entangled States. Int. J. Theor. Phys. \textbf{50}, 1793-1802 (2011)
\bibitem{09_Chong}Chong, S.K., Hwang, T.: Quantum key agreement protocol based on BB84. Opt. Commun. \textbf{283}, 1192-1195 (2010)
\bibitem{010_QKA2012}Shi, R.H.,Zhong, H.: Multi-party quantum key agreement with bell states and bell measurements. Quantum Inf. Process. \textbf{12}, 921-932 (2013)
\bibitem{011_LB2012}Liu, B.,  Gao, F., Huang, W., Wen, Q.Y.: Multiparty quantum key agreement with single particles. Quantum Inf. Process.
\textbf{12}, 1797-1805 (2013)
\bibitem{012_SZW2013}Sun, Z.W., Zhang, C., Wang, B.H., Li, Q., Long, D.Y.: Improvements on ``multiparty quantum key agreement
with single particles". Quantum Inf. Process. DOI: 10.1007/s11128-013-0608-7 (2013)
\bibitem{013-qsj09}Gao, F., Qin, S.J., Guo, F.Z., Wen, Q.Y. : Cryptanalysis of the arbitrated quantum signature protocols.  Phys. Rev. A \textbf{84}, 022344 (2011)
\bibitem{014-qsj09}Lo, H.K., Ko, T.M.: Some attacks on quantum-based cryptographic protocols.  Quantum Inf. Comput. \textbf{5}, 40-47 (2005)
\bibitem{015-GaoPRL}Gao, F.,  Guo, F.Z., Wen, Q.Y.,  Zhu, F.C.: Comment on ``experimental demonstration of a quantum protocol for Byzantine agreement
and liar detection".  Phys. Rev. Lett.  \textbf{101},  208901 (2008)
\bibitem{016-GF07}Gao. F., Wen, Q.Y., Zhu, F.C.: Comment on: ``Quantum exam". Phys. Lett. A \textbf{360}, 748 (2007)
\bibitem{017-qsj10}Qin, S.J., Gao, F., Guo, F.Z.,  Wen, Q.Y.: Comment on ``Two-way protocols for quantum cryptography with a nonmaximally entangled qubit pair". Phys. Rev. A \textbf{82}, 036301 (2010)
\bibitem{018-N.Gisin}Gisin, N., Fasel, S., Kraus, B.,  Zbinden, H., Ribordy, G.: Trojan-horse attacks on quantum-key-distribution systems.  Phys. Rev. A  \textbf{73}, 022320  (2006)
\bibitem{019_GF}Gao, F., Qin, S.J., Wen, Q.Y., Zhu, F.C.: A simple participant attack on the br\'{a}dler-du\v{s}ek protocol. Quant. Inf. Comput. \textbf{7}, 329-334 (2007)
\bibitem{020-HW13}Huang, W.,  Zuo, H.J.,  Li, Y.B.: Cryptanalysis and Improvement of a Multi-User Quantum Communication Network Using ¦Ö-Type Entangled States. Int. J. Theor. Phys. \textbf{52}, 1354-1361 (2013)
\bibitem{021_GF08}Gao, F., Guo, F.Z., Wen, Q.Y., Zhu, F.C.: Comment on ``Experimental Demonstration of a Quantum Protocol for Byzantine Agreement and Liar Detectio''. Phys. Rev. Lett. \textbf{101}, 208901 (2008)
\bibitem{022_Gisin N}Gisin, N., Ribordy, G., Tittel, W., Zbinden, H.: Quantum cryptography. Rev. Mod. Phys. \textbf{74}, 145-195 (2002).
\bibitem{023_Inamori H}Deutsch, D., Ekert, A., Jozsa, R., Macchiavello, C.,  Popescu, S.,  Sanpera, A.: Quantum Privacy Amplification and the Security of Quantum Cryptography over Noisy Channels. Phys. Rev. Lett. \textbf{77}, 2818 (1996).


\end{thebibliography}


\end{document}